# The essential role of multi-point measurements in turbulence investigations: the solar wind beyond single scale and beyond the Taylor Hypothesis

W. H. Matthaeus[1]
Department of Physics and Astronomy, University of Delaware, Newark DE 19716 USA
email: whm@udel.edu  Tel: (302)-983-8767

Co-authors:
R. Bandyopadhyay[1], M.R.Brown[2], J. Borovsky[3], V. Carbone[4], D. Caprioli[5], A. Chasapis[1], R. Chhiber[1], S. Dasso[6,24], P. Dmitruk[6], L.Del Zanna[23], P. A. Dmitruk[6], Luca Franci[7], S.P. Gary[3], M. L. Goldstein[3], D. Gomez[6], A. Greco[4], T.S.Horbury[8], Hantao Ji[26], J.C.Kasper[9], K.G. Klein[10], S. Landi[23], Hui Li[11], F. Malara[4], B. A. Maruca[1], P.Mininni[6], Sean Oughton[12], E. Papini[23], T. N. Parashar[1], Arakel Petrosyan[13], Annick Pouquet[14], A. Retino[25], Owen Roberts[15], David Ruffolo[16], Sergio Servidio[4], Harlan Spence[17], C. W. Smith[17], J. E. Stawarz[8], Jason TenBarge[22], B. J. Vasquez[17], Andris Vaivads[18], F.Valentini[4], Marco Velli[19], A. Verdini[23], Daniel Verscharen[20], Phyllis Whittlesey[21], Robert Wicks[20], R.Bruno[22], G.Zimbardo[4]

[1]University of Delaware, Newark DE USA, [2]Swarthmore College, Swarthmore, PA USA, [3]Space Science Institute, Boulder CO USA, [4]Università della Calabria, Rende, Italy, [5]University of Chicago, USA, [6]University of Buenos Aires, Argentina, [7]Queen Mary University of London, UK, [8]Imperial College London, London, UK, [9]University of Michigan, Ann Arbor MI USA, [10]University of Arizona, USA[11], Los Alamos national Laboratory, Los Alamos NM, USA, [12]University of Waikato, Hamilton NZ, [13]Space Research Institute, Russian Academy of Sciences, Moscow, [14]University of Colorado Boulder USA, [15]Space Research Institute, Austrian Academy of Sciences, Graz, Austria, [16]Mahidol University, Bangkok, Thailand, [17]University of New Hampshire, Durham NH, USA, [18]Swedish Institute of Space Physics, Uppsala, Sweden, [19]University of California Los Angeles, CA USA, [20]University College London, UK, [21]University of California Berkeley, CA USA, [22]Princeton University, NJ USA, [23]University of Florence, Italy, [24] CONICET, Buenos Aires, Argentina, [25]Ecole Polytechnique, Palaiseau, France; [26]Princeton University, USA

*Introduction.* Turbulence refers to complex dynamics of fluid and plasma systems when nonlinear effects such as advection are stronger than dissipative effects. Dimensionless parameters such as a Reynolds number measure the ratio of the strengths of nonlinear and dissipative effects. The usual picture of turbulence begins with a source of large length scale fluctuations which, by means of nonlinear processes, transfers energy by cascade processes across the 'inertial range' to the shorter scale lengths of the 'dissipation range' where the energy is dissipated into thermal motions of the fluid or plasma. Small scale turbulent motions become so disorderly that theory frequently employs statistical descriptions, even if the dynamics is formally deterministic [1]. Revealing the physics of turbulence in the heliosphere (with implications for astrophysical plasmas in general) will require multi-point observations and an array of spacecraft with multiple inter-spacecraft spatial separations [2]. Such a mission in the solar wind is now technologically feasible, and its implementation will have an enormous impact on heliospheric applications [3]

Complex dynamical couplings in turbulence lead to small-scale dissipation of the energy supplied at large scales, a process described as a cascade. Experiments, observations and numerical simulations all show that analogous descriptions apply to hydrodynamic fluids, magnetofluids (MHD), and weakly collisional plasmas. The greatest similarities are found at the larger scales, while plasmas differ at small scales and high frequencies due to the deficit of collisions and the concomitant emergence of complex kinetic physics. Understanding turbulence requires intensive study of statistical properties for the varying parameters found in nature. We argue that multi-point measurements over a range of scales are required to make significant progress in solar wind physics, which remains the only large turbulent space plasma for which such a program is feasible. An array of spatially distributed spacecraft making measurements at moderately high time cadence can provide a wealth of information to inform space physics applications, including space weather [37], and would be of great importance in more distant plasma venues, from the corona to the interstellar medium.

*Turbulence effects in the heliosphere.* The effects of turbulence are intrinsically multi-scale, and the solar wind cascade process spans decades of space and time scales. One might view the turbulence cascade as a primary way in which cross-scale couplings are enabled, connecting macroscopic and microscopic physics in essential ways. Among the most impactful macroscopic influences of heliospheric turbulence [3] is the putative heating of the corona [31] and subsequent acceleration of the solar wind, which in spite of numerous supporting observations remains to be fully established through missions such as Parker Solar Probe and Solar Orbiter. In the solar wind, extended heating is likely also due to turbulent cascade [4] which operates at different rates in the high cross helicity fast wind, and in the lower cross helicity slow wind [5]. Cross helicity (Alfvenicity) slows turbulence initially, but eventually expansion [6] and shear [7] cause systematic reduction of this Alfvenic correlation. Similar turbulence effects account for the radial behavior of the Alfven ratio (or, residual energy), and for spectral steepening in Helios data [8]. Accordingly, turbulence also appears to account well for the radial evolution of the (low-frequency) spectral breakpoint that is closely associated with the systematic increase of the correlation scale of the fluctuations. It is noteworthy that none of these effects are accounted for by the WKB theory of non-interacting waves [3].

All of the above effects are essentially at the larger collective fluid-like scales. Over a range of scales, extending several decades towards the smaller range, theory suggests that the dynamical development of heliospheric turbulence is responsible for the very important observed features of anisotropy [9] and intermittency [10]. See [3] for details.

Another arena in which turbulence is a major player is the transport, scattering, and acceleration of suprathermal and energetic charged particles. In this case, effects such as pitch-angle scattering operate in a truly cross-scale manner, with solar wind thermal protons resonantly interacting with turbulent fluctuations at the scale of a few hundred kilometers at 1au, while 10 Gev cosmic rays or SEPs resonantly interact with fluctuations at scales of millions of kilometers. Turbulence amplitudes and spectral anisotropy are central in controlling interactions, including resonances, with these energetic particles, e.g., [11].

Given all these demonstrated or anticipated influences, one may reasonably ask at what level do we understand the turbulence that produces these diverse effects in the heliosphere? The answer seems to be that, even with numerous accumulated observational constraints and a reasonable level of progress based on simulation and theory, there are fundamental questions that remain to be addressed experimentally. Simple, idealized steady state inertial range phenomenologies can provide motivation for observed spectral slopes, but physical understanding of these diverse cross scale effects, even in the inertial range, requires deeper knowledge and more advanced observations. Beyond inertial range issues, there are questions about dissipation that involve structures and dynamics at sub-proton kinetic scales [12 -14]. Due to the cross-scale couplings and cascade mechanisms involved, the kinetic processes are necessarily driven by the cascade from larger energy-containing scales [15]. This poses further observational challenges for understanding dissipative structures and bulk heating in the corona and solar wind.

*Major questions in solar wind turbulence.* There are numerous outstanding issues about heliospheric turbulence that have not yet been addressed in observations, *in particular due to lack of sufficient spatial and temporal resolution.* Without the associated observations, the field cannot advance beyond its current status. A few examples are given here.

*Unraveling correlations & structure in space and time.* Solar wind researchers are accustomed to employing the Taylor hypothesis, while plasma wave theorists are accustomed to invoking linear dispersion relations. Both of these provide a one-to-one correspondence of variations in space and variations in time. However, in general, spatial and temporal structure are independent entities. For example, the correlation scale is properly defined using single time multi-point measurements [16]. In general, unraveling the space-time relationship is a necessary goal in quantifying and distinguishing effects of turbulence, waves, reconnection and other phenomena in space plasmas. **Revealing how turbulent energy in a space plasma is distributed in space & time requires multi-point measurements.**

*Anisotropy of the spectrum at varying scales.* What are the cascade rate and the heating rate? Can turbulence explain observed heating and the origin of the solar wind? Spectral information relative to preferred directions, e.g., radial and magnetic field directions, is required to validate or controvert available theoretical explanations. Purely phenomenological treatments do not provide strong conclusions. Anisotropic measurements are required, necessitating **simultaneous multi-point measurements that span 3D spatial directions.**

*Direct measurement of scale transfer.* The Yaglom - Kolmogorov 3rd order laws [17-19] provide a direct evaluation of energy transfer rates at a given scale. The simplest forms require isotropy or some other simple symmetry. Anisotropic forms of the 3rd order law have been applied using Cluster or MMS at single scales, but understanding cross-scale transfer requires anisotropic measurement at several scales. **Simultaneous 3D multi-point measurements are needed to reveal how turbulent energy is transferred anisotropically across scale.**

*Higher order statistics and coherent structures.* Intermittency or patchiness is an essential feature of turbulent heating and cascade processes. Indeed, in strong turbulence at high Reynolds numbers, most statistical measures of spectral transfer and dissipation are highly nonuniform. The fourth order (single time) statistics that provide a baseline measurement of intermittency. The sixth order statistics are a natural measure of patchiness of energy transfer. These are fundamental but have not been fully characterized measured in the solar wind, as they must be measured in anisotropic form, due to the strong influence of the large scale magnetic field and strong gradients, e.g., in stream interaction regions and shear layers [32], as well as in regions of interaction of turbulence with waves [33, 34]. Measurement of 3D structure and orientation of coherent structures near the kinetic proton scales is needed to reveal the role of higher order moments in dissipation, **thus requiring simultaneous 3D multi-point measurements at two or more spatial scales.**

*Anisotropic scale dependent relaxation times.* Because of the classic "closure problem" (2nd moment depends on 3rd, 3rd depends on 4th, etc), higher order statistics at least up to 4th order contain fundamental information about dynamics. The single time statistics are important, but so too are the decay rates of higher order correlations. For example, the decay time of the 3rd order correlations controls spectral evolution. In the context of closures [20], the decay times of the triple correlations are identified with scale-dependent Lagrangian correlation times, and are usually treated as the local Kolmogorov time scale, because the dominant sweeping timescale does not induce spectral transfer. In plasma there are additional available time scales, and understanding 3rd and higher order correlations becomes more complex. Observational constraints, including measurement of anisotropic 4th order (and higher) moments are needed to understand this basic physics. **Multi-spacecraft measurements are required over a wide range of scales to assess these crucial dynamical time scales.**

*Key Turbulence measurements:* A central quantity of interest is the two-point, two-time correlation of a primitive variable (e.g., a magnetic field component b.). This *4D* space-time correlation may be define as $R_{ij}(r,\tau) = \langle b_i(x,t) b_j(x+r, t+\tau) \rangle$ where the brackets denote an ensemble average, or a suitable space-time average. The (trace) wave vector spectrum is $S(k) = [\frac{1}{2\pi}]^3 \int d^3 r R(r,0)\, e^{ik\cdot r}$ in which the time lag is zero, as well as the Eulerian frequency spectrum $E(\omega) = \frac{1}{2\pi} \int d\tau\, R(0,\tau) e^{i\omega\tau}$ in which the spatial lag is zero. The full space-time (trace) spectrum $S(k,\omega)$ is analogously defined as the 4D space-time transform of $R$. This is an analog of a dispersion relation, but without the expectation of definite relationship between frequency and wavevector. If a nonzero time lag is retained when the spatial transform is carried out, one arrives at the important quantity $S(k,\tau) = S(k)\,\Gamma(k,\tau))$. This defines the scale dependent time correlation (in the Eulerian frame) $\Gamma(k,\tau)$, alluded to in the prior section. The space-time correlation also permits direct test of the Taylor hypothesis. Observational determination of the 2nd order, two-time, two-point correlation contains much information that we require, but this is not all that is needed to describe interplanetary turbulence.

**A relation of central importance is the third order law, which directly measures energy transfer across scales.** The contribution of incompressive transfer is given by the Politano-Pouquet law [17]. Hall effect contributions and compressive contributions [21] can be treated additively. With suitable conditions on stationarity, etc., the relevant incompressive differential form is $\nabla_s \cdot \langle \delta z_s^\pm |\delta z_s^\mp|^2 \rangle = -4\epsilon^\mp$ for the ± Elssaser field increments and lag *s*. Integrating over a volume and employing Gauss's law yields a surface integral that determines the total incompressive transfer of the ± fields across that surface. A suitable multi-spacecraft configuration (say, a regular tetrahedron) enables an approximate evaluation of this transfer [18, 22]. *Carrying out this multi-spacraft measurement provides a direct evaluation of scale transfer with no approximations on rotational symmetry.* In various combinations, this approach can be supplemented with, or compared to, single spacecraft results using the frozen-in flow (Taylor) hypothesis, including methods that assume isotropy and other symmetries, e.g., [23]. This approach can also reveal potential cascade to both large and small scales [33,34,35].

**Fourth order correlations are also crucial, as they quantify intermittency, and drive the all-important third order correlations.** In MHD the effect of a mean magnetic field appears in the moment hierarchy at the same order as the 4$^{th}$ order correlations [36]. Together with the mean field, the 3rd and 4th order correlations influence the production of spectral anisotropy [9, 24], a major issue in plasma cascade and dissipation [14]. Also at 4$^{th}$ order is the anisotropic *scale dependent kurtosis,* a quantity that reveals scale-varying intermittency, anisotropy of coherent structures, and incoherent wave activity, as seen in the examples [25] using MMS data.

Much theoretical attention is paid to the inertial and kinetic cascade ranges in plasma turbulence, but the last stages in which collective flow and field energies are converted into microscopic motions or "heat" are crucial for understanding dissipation. Two quantities of great importance in this regard are the work done on particles of species α, that is, $J^\alpha \cdot E$ by the e/m fields, where $E$ is the total electric field, and the pressure strain interaction $\Pi_{ij}^\alpha S_{ij}^\alpha$ where $\Pi^\alpha$ is the pressure tensor and $S = \partial_i u_j^\alpha + \partial_j u_i^\alpha$ is the symmetric rate of strain tensor, each of species α [26, 27]. Even though these quantities are not sign-definite, as viscous dissipation would be, their net (averaged) values are interpreted as the conversion of e/m energy into flow energy, and the conversion of gradients in the flow into internal energy. These channels of energy conversion are agnostic regarding specific mechanisms (e.g., reconnection) that may be producing heating, and are therefore crucial diagnostics for understanding the termination of the cascade and the degeneration of collective motions into internal energy. Multi-spacecraft techniques again enter prominently, as the total $J$ can be evaluated by curlometer techniques, while the rate of strain tensor can be similarly evaluated by differencing the velocities across spacecraft pairs.

*Conclusion.* Statistical quantities such as those above are essential to full understanding of turbulence, cascade and dissipation in a magnetized plasma. Correlations are expected to be anisotropic and proper analysis requires measurement at several lag scales. Reasonable choices are, first, near the ion kinetic scales, and then at two or more larger scales, where the fluid MHD-like cascade remains operative. Methods have been developed to extract space-time information from multi-spacecraft datasets, including wave telescope (or k-filtering) [28] and direct methods that rely on ensemble statistics [29]. Some such methods have been successfully applied in plasma laboratory experiments [e.g., 38]. *When a mission with a sufficient number of spacecraft* *is deployed in the solar wind, multi-point methods will answer many basic questions about interplanetary and astrophysical turbulence that can be experimentally addressed in no other way [2].* **Multi-point, multi-scale in-situ measurement, as discussed for a HelioSwarm mission, for example, is a unique approach to revealing the fundamental nature of turbulence in the solar wind [30], with immediate implications for the corona [31] and other space and astrophysical plasmas.**